\documentclass[aps,twocolumn,superscriptaddress,prl]{revtex4-1}

\usepackage{bm,graphicx,textcomp,amssymb,amsmath,dcolumn,color}

\newcommand{\tr}{\operatorname{tr}}

\renewcommand{\Re}{\operatorname{Re}}
\renewcommand{\Im}{\operatorname{Im}}

\newcommand{\ii}{{\rm i}}
\newcommand{\ket}[1]{\left|#1\right>}
\newcommand{\bra}[1]{\left<#1\right|}
\newcommand{\abs}[1]{{\left|#1\right|}}
\newcommand{\braket}[2]{\left<#1|#2\right>}
\newcommand{\nn}{\nonumber\\}

\newcommand{\f}[1]{\mbox{\boldmath$#1$}}

\newcommand{\na}{\mbox{\boldmath$\nabla$}}
\newcommand{\bea}{\begin{eqnarray}}
\newcommand{\ea}{\end{eqnarray}}
\newcommand{\eea}{\end{eqnarray}}
\newcommand{\ord}{\,{\cal O}}

\newcommand{\expval}[1]{\langle #1\rangle}

\newcommand{\new}[1]{{
#1}}

\begin{document}

\title{Quantum Zeno \new{Manipulation} of Quantum Dots}

\author{N.~Ahmadiniaz} 

\affiliation{Helmholtz-Zentrum Dresden-Rossendorf, 
Bautzner Landstra{\ss}e 400, 01328 Dresden, Germany,}

\author{M.~Geller} 

\author{J.~K\"onig} 

\author{P.~Kratzer} 

\author{A.~Lorke} 

\affiliation{Fakult\"at f\"ur Physik and CENIDE, Universit\"at Duisburg-Essen,
  Lotharstra{\ss}e 1, 47057 Duisburg, Germany,}



\author{G.~Schaller}

\affiliation{Helmholtz-Zentrum Dresden-Rossendorf, 
Bautzner Landstra{\ss}e 400, 01328 Dresden, Germany,}

\author{R.~Sch\"utzhold}

\affiliation{Helmholtz-Zentrum Dresden-Rossendorf, 
Bautzner Landstra{\ss}e 400, 01328 Dresden, Germany,}

\affiliation{Institut f\"ur Theoretische Physik, 
Technische Universit\"at Dresden, 01062 Dresden, Germany,}

\date{\today}

\begin{abstract}
We investigate whether and how the quantum Zeno effect, i.e., the inhibition 
of quantum evolution by frequent measurements, can be employed to isolate 
a quantum dot from its surrounding electron reservoir. 
In contrast to the often studied case of tunneling between discrete levels, 
we consider the tunnelling of an electron from a continuum reservoir to a 
discrete level in the dot.
%
Realizing the quantum Zeno effect in this scenario can be much harder because 
the measurements should be repeated before the wave packet of the hole left 
behind in the reservoir moves away from the vicinity of the dot. 
Thus, the required repetition rate could be lowered by having a flat band 
(with a slow group velocity) in resonance with the dot or a sufficiently 
small Fermi velocity or a strong external magnetic field.
\new{We also consider} 
\new{the anti-Zeno effect, i.e., how measurements can 
accelerate or enable quantum evolution.}
\end{abstract}

\maketitle

\paragraph{Introduction} 

One of the major distinctions between classical and quantum physics is the role 
of measurements.
As a consequence, it is impossible to directly observe quantum evolution taking 
place without actually affecting it. 
A striking example is the quantum Zeno effect which describes slowing down or 
even stopping quantum evolution by frequent measurements, see, 
e.g.~\cite{misra1977a,itano1990a,hackenbroich1998a,fischer2001a,streed2006a,bernu2008a,slichter2016a}.

Let us discuss the basic picture by means of a simple example, two discrete 
states or levels $\ket{1}$ and $\ket{2}$ of equal energy which are 
tunnel-coupled.
Preparing the initial state in one level $\ket{\Psi(t=0)}=\ket{1}$, 
the time-dependent probability $P_2(t)=|\braket{2}{\Psi(t)}|^2$ 
for the other level reads $P_2(t)=\sin^2(\gamma t)$, 
where the frequency $\gamma>0$ is given by the tunneling strength.
For short times $t\ll1/\gamma$, we obtain a quadratic growth 
$P_2(t)\approx\gamma^2t^2$, because in quantum physics amplitudes instead 
of probabilities are added. 
However, if we measure the level occupation after such a short time, i.e., 
within the quadratic-growth regime, we project the state $\ket{\Psi(t)}$
back to the initial state $\ket{1}$ 
(assuming a strong, i.e., projective measurement) 
with high probability $P_1(t)\approx1-\gamma^2t^2$ 
and thus the quantum evolution has to start again afterwards. 
In other words, the amplitudes before and after the measurement do no longer 
interfere constructively because we have obtained which-way information via 
the measurement. 
Now, repeating this measurement with a fast rate (much larger than $\gamma$) 
would effectively keep setting back the quantum 
evolution such that the quantum state stays in the initial level $\ket{1}$.
This inhibition (or slowing down) of quantum evolution is usually referred 
to as the quantum Zeno effect~\cite{misra1977a}.
Note that this line of argument crucially relies on the quadratic growth 
$P_2(t)\approx\gamma^2t^2$ discussed above. 
In case of a linear behavior $P_2(t)\sim t$, for example, setting back the 
evolution by measurements would not have this retarding effect. 

The quantum Zeno effect is a striking example for quantum 
control~\cite{facchi2001a,facchi2002a,facchi2004a,wiseman2010,alvarez2010a,paz_silva2012a,schaefer2014a,barontini2015a} and 
closely related to passive quantum error correction schemes 
(similar to the spin-echo method). 
Thus, this fascinating phenomenon is of fundamental interest and has 
already been observed experimentally in different systems, e.g., 
for ions in Penning traps~\cite{itano1990a}, 
ultra-cold atoms in optical lattices~\cite{fischer2001a,zhu2014a}, 
Bose-Einstein condensates in magnetic traps~\cite{streed2006a,schaefer2014a}, 
Rydberg atoms in cavities~\cite{bernu2008a} 
or superconducting qubits~\cite{slichter2016a}.

On much shorter time scales, electrons in quantum dots are also very 
suitable systems for studying the quantum Zeno effect:  
Their energy levels and tunneling rates can be manipulated  
(e.g., by the variation of voltages that are applied to suitably 
placed gate electrodes) 
or the size of the dot and they 
have quite long coherence times while their states can be 
read out, i.e., measured, quite fast by capacitive charge 
detectors~\cite{fujisawa2006a,flindt2009a} or optical 
transitions~\cite{kurzmann2016a,kurzmann2019a}, 
for example. 

In most previous works, the quantum Zeno effect has been studied
in a regime where the aforementioned picture based on transitions 
between discrete levels can be 
applied~\cite{itano1990a,hackenbroich1998a,gurvitz2003a,gurvitz2006a,segal2007a,schaller2018a,leppenen2022a}. 
%
In the following, we shall study the more involved case of transitions 
between a discrete level and a continuum (a Fermi gas or liquid), 
see also e.g.~\cite{gurvitz1997a,engelhardt2018a}.
As a concrete experimental realization, we consider a quantum dot coupled 
to a reservoir in the form of an effectively two-dimensional electron gas 
(2DEG)~\cite{ando1982a,reimann2002b,russ2004a,russ2006a,geller2019a}.  


\paragraph{The Model} 

\begin{figure}
\includegraphics[width=0.3\textwidth]{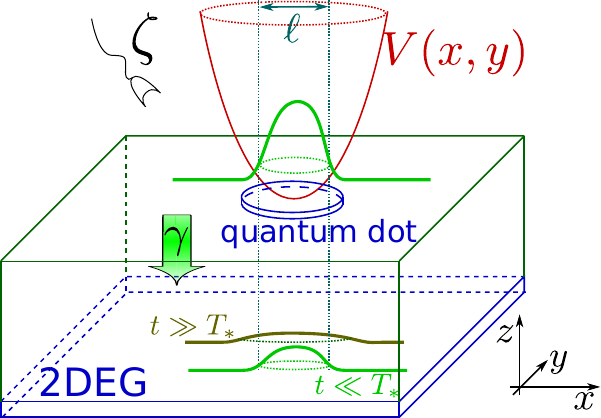}
\caption{\label{FIG:sketch}
Sketch of the quantum dot with confinement potential $V(x,y)$
coupled to the effectively two-dimensional electron gas (2DEG), 
\new{both realized as lowest quantum well states}.  
%
Tunneling with strength $\gamma$ of an electron or hole wave-packet 
with characteristic size $\ell$ from the dot to the 2DEG reservoir 
corresponds to discharging or charging.
In the reservoir, the wave-packets then spread or move away on a typical 
time scale $T_*$ which limits 
the repetition rate of the measurement 
(indicated by $\zeta$) required for observing the quantum Zeno effect. 
%
}
\end{figure}

We consider the quantum dot-reservoir system in good 
approximation~\cite{vdovin2000a} being described
by the following many-body Hamiltonian ($\hbar=1$), cf.~Fig.~\ref{FIG:sketch}
\bea
\label{model}
\hat H
&=&
\int d^2r
\left(
\frac{(\na\hat\psi_1^\dagger)\cdot(\na\hat\psi_1)}{2m}
+V_1(x,y) \hat\psi_1^\dagger\hat\psi_1+
\right.
\nn 
&&
+
\left.
\frac{(\na\hat\psi_2^\dagger)\cdot(\na\hat\psi_2)}{2m}
+\left(\gamma\hat\psi_1^\dagger\hat\psi_2+{\rm h.c.}\right) 
\right) 
\,.
\ea
The first line represents the quantum dot for which we employ the 
standard~\cite{warburton1998a} harmonic potential approximation 
$V_1(x,y)=V_0+m\omega^2(x^2+y^2)/2$ 
where the offset $V_0$ can be tuned by a gate voltage. 
The second line describes the reservoir and its tunnel coupling to the 
dot with the coupling strength $\gamma$.
The field operators $\hat\psi_1(x,y)$ and $\hat\psi_2(x,y)$ could be 
envisaged as lowest quantum well states in $z$-direction, 
for example, 
\new{where we assume that the energies of the higher quantum well states 
are so large (tight confinement limit) that we can neglect them.} 

If we focus on the ground state of the quantum dot 
$\hat\psi_1(x,y)\to\Phi_{0}(x,y)\hat a_{\rm d}$ and Fourier transform the 
reservoir modes 
$\hat\psi_2(x,y)\to\hat\psi_\mathbf{k}$, the Hamiltonian~\eqref{model}
becomes 
\bea
\label{Fourier}
\hat H
=
\epsilon_{\rm d}\hat n_{\rm d} 
+
\int d^2k\left[\epsilon_\mathbf{k}\hat\psi_\mathbf{k}^\dagger\hat\psi_\mathbf{k}
+\left(
\gamma_\mathbf{k}\hat a_{\rm d}^\dagger\hat\psi_\mathbf{k}+{\rm h.c.}
\right)
\right] 
\,,
\ea
where $\epsilon_{\rm d}$ denotes the energy of the quantum dot level and 
$\hat n_{\rm d}=\hat a_{\rm d}^\dagger\hat a_{\rm d}$ the 
corresponding number operator, while 
$\epsilon_\mathbf{k}=\mathbf{k}^2/(2m)$
are the single-particle energies of the reservoir modes $\mathbf{k}=(k_x,k_y)$. 
Finally, $\gamma_\mathbf{k}$ is determined by the Fourier transform of the 
ground-state wave-function inside the quantum dot 
(see e.g.~\cite{wibbelhoff2005a}), i.e., Gaussian 
\bea
\label{Gaussian}
\gamma_\mathbf{k}=\gamma \int \frac{d^2r}{2\pi} \Phi_{0}^*(x,y) 
e^{+\ii \mathbf{k}\cdot\mathbf{r}}
=\gamma_0\exp\left\{-\frac12\mathbf{k}^2\ell^2\right\}
\,,
\ea
where $\ell=\sqrt{\hbar/(m \omega)}$ is the characteristic dot length scale.
Note that the bi-linear Hamiltonian~\eqref{model} does not explicitly include 
non-linear (e.g., Coulomb) interactions between the electrons nor their 
coupling to other degrees of freedom (e.g., phonons) whose potential 
impact will be discussed below.
Therefore, we restrict ourselves to studying the transition between an empty 
and a singly-occupied dot, for which a charging energy of the quantum dot is 
irrelevant.

\paragraph{Quantum Zeno Dynamics} 

Now we are in the position to study the tunneling process of an electron from 
the reservoir into the quantum dot. 
We start with the initial state $\ket{\rm in}$ where the dot is empty 
$\hat a_{\rm d}\ket{\rm in}=0$ while the reservoir is filled up to the 
Fermi energy. 

Although the above model can be solved exactly using, e.g., Greens function 
techniques~\cite{haug2008}, 
we may already calculate the time-dependent probability $P(t)$ for the occupation 
of the quantum dot using
standard time-dependent perturbation theory with respect to $\gamma$,
\bea
\label{sinc}
P(t)
=\expval{\hat n_{\rm d}(t)}
=4\int\limits_{\rm F} d^2k\;|\gamma_\mathbf{k}|^2
\frac{\sin^2([\epsilon_{\rm d}-\epsilon_\mathbf{k}]t/2)}
{(\epsilon_{\rm d}-\epsilon_\mathbf{k})^2}
\,,
\ea
where the subscript F on the integral indicates that the integration runs 
up to Fermi energy $\epsilon_{\rm F}$.

\paragraph{Quantum-Zeno Regime} 

Even within the region of applicability of perturbation theory 
(i.e., for small probabilities $P(t)\ll1$), we obtain two temporal regimes. 
For small times $t$, the probability $P(t)$ grows quadratically 
\bea
\label{quadratic}
P(t)=t^2\int\limits_{\rm F} d^2k\;|\gamma_\mathbf{k}|^2+\ord(t^3)
\,.
\ea
For the Gaussian case~\eqref{Gaussian}, this simplifies to 
\bea
\label{quadratic-Gaussian}
P(t)=
\pi\,\frac{\gamma_0^2t^2}{\ell^2}\left(1-\exp\left\{-k^2_{\rm F}\ell^2\right\}
\right) 
+\ord(t^3)
\,,
\ea
where $k_{\rm F}=\sqrt{2m\epsilon_{\rm F}}$ denotes the Fermi momentum. 
If the Fermi energy $\epsilon_{\rm F}$ is large enough
$2m\epsilon_{\rm F}\ell^2\gg1$, i.e., $k^2_{\rm F}\ell^2\gg1$, 
we obtain $P(t)=\pi\gamma_0^2t^2/\ell^2$.  
In the opposite limit $k^2_{\rm F}\ell^2\ll1$, the integration
is cut off by $k^2_{\rm F}$ and we find $P(t)=\pi\gamma_0^2t^2k^2_{\rm F}$.

\paragraph{Regime of Fermi's Golden Rule} 

For later times (but still in the perturbative regime $P(t) \ll 1$), 
we can use the standard approximation of the sinc function in 
Eq.~\eqref{sinc} by a Dirac delta function and obtain a linear growth in time, 
consistent with Fermi's Golden Rule 
\bea
\label{golden}
P(t)\approx2\pi t\int\limits_{\rm F} d^2k\;|\gamma_\mathbf{k}|^2
\delta(\epsilon_{\rm d}-\epsilon_\mathbf{k})
\,.
\ea
Obviously, the above integral vanishes for $\epsilon_{\rm d}>\epsilon_{\rm F}$
and for $\epsilon_{\rm d}<0$, i.e., if there are no filled reservoir modes in 
resonance with the dot. 
\new{(These cases are discussed in the supplement.)}
%
%
Assuming $\epsilon_{\rm F}>\epsilon_{\rm d}>0$, we find for the Gaussian 
case~\eqref{Gaussian} with $\epsilon_\mathbf{k}=\mathbf{k}^2/(2m)$
\bea\label{linear-Gaussian}
P(t)\approx(2\pi)^2tm\gamma_0^2\exp\left\{-2m\epsilon_{\rm d}\ell^2\right\}
\,.
\ea
%
The exponential suppression $\exp\left\{-2m\epsilon_{\rm d}\ell^2\right\}$
for large dot energies $2m\epsilon_{\rm d}\ell^2\gg1$ stems from the
wave-function mismatch between the Gaussian ground-state wave-function of 
the dot and the reservoir wave-functions of the corresponding energy 
$\epsilon_{\rm d}$ which are rapidly oscillating for large $\epsilon_{\rm d}$.
In order to avoid this suppression, we assume small dot energies 
$2m\epsilon_{\rm d}\ell^2\ll1$ in the following. 

As a peculiarity of the case of two spatial dimensions, the remaining 
pre-factor $P(t)\approx(2\pi)^2tm\gamma_0^2$ is actually independent of 
$\epsilon_{\rm d}$ because the energy $\epsilon_\mathbf{k}=\mathbf{k}^2/(2m)$
and the volume factor $d^2k$ both display the same quadratic scaling in $k$,
such that the density of states per energy is constant.

\paragraph{Crossover Time} 

Having found an initial period of quadratic growth~\eqref{quadratic} for early 
times (quantum Zeno regime) followed by a period of linear 
growth~\eqref{golden} for later times (regime of Fermi's Golden Rule), 
we may estimate the crossover time $T_*$ marking the transition between 
the two regimes~\cite{sangeetha} by comparing Eqs.~\eqref{quadratic} 
and~\eqref{golden}.
For the Gaussian limit we obtain with~\eqref{quadratic-Gaussian} 
and~\eqref{linear-Gaussian}
\bea
T_*\approx
4\pi m\ell^2
\frac{\exp\left\{-2m\epsilon_{\rm d}\ell^2\right\}}
{1-\exp\left\{-\ell^2k^2_{\rm F}\right\}}
\,.
\ea
Assuming small dot excitation energies $2m\epsilon_{\rm d}\ell^2\ll1$ and 
considering the limiting cases as discussed after 
Eq.~\eqref{quadratic-Gaussian}, this simplifies to 
$T_*\approx4\pi m\ell^2$ for $\ell^2k^2_{\rm F}\gg1$ and to 
$T_*\approx4\pi m/k^2_{\rm F}$ for $\ell^2k^2_{\rm F}\ll1$, 
respectively. 

As explained in the Introduction, only measurements with a repetition
time faster than $T_*$ can induce the quantum Zeno effect.
This time scale is typically quite short and can be visualized by the 
following intuitive picture.

Let us first consider the case $\ell^2 k_{\rm F}^2\gg1$.
If an electron tunnels with a small amplitude from the reservoir to 
the quantum dot, it leaves behind a hole in the reservoir. 
The typical size of the initial wave packet of this hole is determined 
by the characteristic length $\ell$ of the ground-state wave function 
of the dot.
Afterwards, this wave packet is spreading out or moving away with the 
typical group velocity at that length scale $v_{\rm group}\propto1/(\ell m)$.
Once the wave packet has spread out too much or moved away too far, 
further tunneling amplitudes would not be added coherently and 
probabilities would add up instead. 
This is precisely the transition from the quantum Zeno regime to the 
regime of Fermi's Golden Rule occurring at the crossover time 
$T_*\sim\ell/v_{\rm group}$.

In the opposite limit ($k_{\rm F}^2 \ell^2 \ll 1$), one should just replace the 
length scale $\ell$ of the dot by the Fermi length 
$\ell_{\rm F}\propto1/k_{\rm F}$ and the associated group velocity by 
the Fermi velocity $v_{\rm F}=k_{\rm F}/m$. 
%
Then, the crossover time $T_*$ is basically the Fermi time $T_{\rm F}$ 
which is related to the Fermi energy $\epsilon_{\rm F}$ by Heisenberg's 
uncertainty relation [$T_* =\ord\{\epsilon_{\rm F}^{-1}\}$] 
and determines the minimum response time of hole
states in the reservoir. 

\new{Consistent with the results of~\cite{gurvitz1991a}, an effectively Markovian 
reservoir is obtained in the combined limit of $\ell\to0$ and 
$k_{\rm F}\to\infty$ 
where the available density of states in the reservoir becomes constant 
and the quantum-Zeno regime/effect disappears $T_*=0$.}

\paragraph{General Band Structure} 

Let us now discuss possible generalizations of the above results 
{(see also the supplement).}
For reservoir electrons propagating in a lattice, the single-particle 
energies $\epsilon_\mathbf{k}$ of the reservoir modes $\mathbf{k}$ 
should be replaced by the lattice band structure, which might also 
modify the (experimentally accessible~\cite{zhou2015a}) couplings 
$\gamma_\mathbf{k}$ 
accordingly. 
Of course, the same would happen for non-harmonic quantum dot potentials 
$V_1(x,y)\neq V_0+m\omega^2(x^2+y^2)/2$.
Nevertheless, one would still expect the effective 
Hamiltonian~\eqref{Fourier} to provide a good approximation. 
Thus, the results~\eqref{sinc} and thus \eqref{quadratic} and 
\eqref{golden} remain valid, but with modified $\epsilon_\mathbf{k}$
and $\gamma_\mathbf{k}$.
This may entail interesting consequences.
For example, if we have a flat band 
(with $\epsilon_\mathbf{k}=\rm const$)
below the Fermi energy in resonance with the quantum dot 
(i.e., $\epsilon_\mathbf{k}=\rm const=\epsilon_{\rm d}$),
the probability~\eqref{golden} per time would become very 
large while the initial growth~\eqref{quadratic} would remain 
almost unaffected since it is independent of $\epsilon_\mathbf{k}$.
This means that the crossover time $T_*$ would also become very 
large, which might help observing the quantum Zeno effect in this 
scenario. 
In terms of the intuitive picture sketched above, the group velocity 
becomes very small for such a flat band such that the wave packet stays 
quite long in the vicinity of the dot. 

\paragraph{Magnetic Field} 

The above finding regarding flat or nearly flat bands motivates the study of 
a strong external magnetic field perpendicular to the reservoir, because this 
does also turn the reservoir modes into flat bands in the form the well-known 
Landau levels. 

As usual, we represent the constant perpendicular magnetic field 
$\mathbf{B}=B\mathbf{e}_z$ in the symmetric gauge by the vector potential 
$\mathbf{A}=\mathbf{B}\times\mathbf{r}/2$.
Then, after minimal coupling $\na\to\na-q\mathbf{A}$ to the electron charge
$q$, the Hamiltonian~\eqref{model} acquires an additional angular-momentum 
contribution $-2\mathbf{A}\cdot\na=\mathbf{B}\cdot\hat{\mathbf{L}}=B\hat L_z$ 
as well as a harmonic confinement potential 
$V_B=\Omega_B^2(x^2+y^2)/8$ from the quadratic term 
$\mathbf{A}^2=B^2(x^2+y^2)/4$ where $\Omega_B$ is the cyclotron frequency 
$\Omega_B=qB/m$. 

Thus, the reservoir eigen-functions are no longer plane waves as in 
Eq.~\eqref{Fourier}, but discrete Landau levels which can be represented 
by the modes of a two-dimensional harmonic oscillator in polar 
coordinates $r$ and $\varphi$~\cite{ciftja2020a}
\bea
\psi_{n,l}(r,\varphi)=f_{n,l}(r) e^{il\varphi} 
\,.
\ea
Here $n\in\mathbb N$ and $l\in\mathbb Z$ are the quantum numbers 
corresponding to energy $\epsilon_n=\Omega_B(n+1/2)$
%
%
and angular momentum $\hat L_z$, respectively. 
The radial mode functions $f_{n,l}(r)$ are given by polynomials 
multiplied by a Gaussian $\exp\{-r^2/(2\ell_B)^2\}$ with the 
magnetic length $\ell_B=1/\sqrt{qB}$. 
For example, the lowest Landau levels with $n=0$ have 
$f_{n=0,l}(r)\propto r^l \exp\{-r^2/(2\ell_B)^2\}$ with $l\geq0$, 
which can be experimentally mapped~\cite{russ2004a,lei2010a}.

The ground-state wave function of the quantum dot will also be 
modified, but merely by the additional harmonic confinement potential 
$V_B=\Omega_B^2(x^2+y^2)/8$ which narrows the Gaussian wave-function
(i.e., decreases $\ell$) and increases the energy $\epsilon_{\rm d}$. 
%
Since this wave function is rotationally symmetric 
(i.e., has zero angular momentum), 
it only tunnel couples to reservoir modes with $l=0$.

If we now assume a strong magnetic field $B$ and/or a small Fermi energy 
$\epsilon_{\rm F}<3\Omega_B/2$ 
such that only the lowest Landau levels 
are occupied, the effective Hamiltonian~\eqref{Fourier} 
can be restricted to the mode $n=l=0$ and reads 
\bea
\label{LLL}
\hat H
=
\epsilon_{\rm d}\hat n_{\rm d}+\frac{\Omega_B}{2}\hat n_{0,0}+ 
\left(
\gamma_{0,0}\hat a_{\rm d}^\dagger\hat a_{0,0}+\rm h.c.
\right)
\,,
\ea
where the effective tunnel coupling $\gamma_{0,0}$
is determined by the overlap 
of the two Gaussian wave functions of the dot and 
$\psi_{0,0}\propto\exp\{-r^2/(2\ell_B)^2\}$.

Assuming resonance $\epsilon_{\rm d}=\Omega_B/2$, this 
Hamiltonian~\eqref{LLL} is then formally equivalent to 
tunneling between two discrete modes 
(described by $\hat a_{\rm d}$ and $\hat a_{0,0}$) such that 
we are back to the original picture of the quantum Zeno effect 
as described in the Introduction. 
\new{Off-resonant scenarios $\epsilon_{\rm d}\neq\Omega_B/2$ 
may give rise to the anti-Zeno effect~\cite{kofmann2000a,facchi2001a,fischer2001a,segal2007a}, as discussed in the supplement.}

\paragraph{Measurement Model} 

So far, the Hamiltonian~\eqref{model} does only describe the internal dynamics 
of the electrons, but not the actual measurement process -- which is causing 
the quantum Zeno effect. 
To incorporate this, let us construct a simple toy model for the read out. 
We assume that the dot is strongly illuminated by laser light (saturation regime) 
which quickly transfers the electron in the quantum dot from the ground state 
level to an excited level $\hat a_{\rm d} \hat a_{\rm e}^\dagger$ as soon as 
the dot is occupied. 
%
This excited level then rapidly decays by emitting a photon as 
described by the bosonic creation and annihilation operators $\hat b^\dagger$ 
and $\hat b$, where we focus on one mode for simplicity. 
Assuming that these excitation--de-excitation cycles occur much faster than 
all the other time scales considered above (e.g., $T_*$), we may use an 
effective description where the dot emits photons as soon as it is occupied.
Thus, we model the measurement by the additional Hamiltonian 
\bea\label{EQ:pointer}
\hat H_{\rm measure}=\hat n_{\rm d}\left(i\zeta\hat b^\dagger+{\rm h.c.}\right) 
\ea
with the detector coupling strength $\zeta>0$, 
which is basically a pointer Hamiltonian for measuring the observable 
$\hat n_{\rm d}$. 

Since $\hat H_{\rm measure}$ commutes with the undisturbed Hamiltonian 
$\hat H_0=\epsilon_{\rm d}\hat n_{\rm d} 
+\int d^2k\,\epsilon_\mathbf{k}\hat\psi_\mathbf{k}^\dagger\hat\psi_\mathbf{k}$, 
we may incorporate it easily into the time-dependent perturbation theory
used above for calculating $P(t)$. 
In the interaction picture, the dot annihilation operator $\hat a_{\rm d}$ 
just acquires an additional operator-valued factor 
\bea\label{EQ:intpic_ms}
\hat a_{\rm d}(t)
=
e^{-i\epsilon_{\rm d}t}
\exp\left\{(\zeta\hat b^\dagger-{\rm h.c.})t\right\}
\hat a_{\rm d}(0)
\,,
\ea
which represents the dynamics induced by {$\hat H_0+\hat H_{\rm measure}$.}
When acting on the initial photonic vacuum state $\ket{0}$ with 
$\hat b\ket{0}=0$,
this additional factor $\exp\{(\zeta\hat b^\dagger-{\rm h.c.})t\}$ just 
generates a coherent state $\ket{\alpha(t)}$ of the photon field whose amplitude 
$\alpha(t)=\zeta t$ grows linearly with time $t$. 

The overlap of these coherent states at different times 
$\braket{\alpha(t_1)}{\alpha(t_2)}=e^{-\zeta^2(t_1-t_2)^2/2}$ 
modifies Eq.~\eqref{sinc} via 
\bea\label{EQ:pocc_withdet}
P(t)
&=&
\int\limits_0^t dt_1\int\limits_0^t dt_2
\int\limits_{\rm F} d^2k\;|\gamma_\mathbf{k}|^2
\exp\left\{-\frac{\zeta^2}{2}(t_1-t_2)^2\right\}
\nn&&\times
\exp\{-i(\epsilon_{\rm d}-\epsilon_\mathbf{k})(t_1-t_2)\}
\,.
\ea
For very short times $t\ll T_*$ and $t\ll1/\zeta$, we recover the 
initial quadratic growth in Eq.~\eqref{quadratic}.
What happens after that initial period depends on the quantity $\zeta T_*$. 
For small $\zeta T_*\ll1$, we recover Eq.~\eqref{golden}. 
For large $\zeta T_*\gg1$, on the other hand, the probability is strongly 
suppressed 
\bea 
P(t)\sim t\,\frac{\sqrt{2}}{\zeta}
\int\limits_{\rm F} d^2k\;|\gamma_\mathbf{k}|^2
\,,
\ea
which is a manifestation of the quantum Zeno effect by frequent measurements 
with the effective rate $\zeta$.

The detector signal can be very explicitly evaluated with methods of 
Full Counting Statistics or quantum polyspectra~\cite{schaller2018a,sifft2021a}.
However, already the analysis of average values reveals (see supplement) 
that -- although delay effects occur -- a transition between the timescales 
can be directly resolved.

\paragraph{Conclusions} 

For the example of a quantum dot tunnel coupled to an effectively 
two-dimensional electron reservoir, we study whether and how the quantum Zeno 
effect can be employed to suppress tunneling between a discrete level and a 
continuum. 
In contrast to tunneling between two discrete levels, we find that the required 
measurement repetition rate is set by the time $T_*$ it takes the wave-packet 
in the reservoir to move away from the vicinity of the quantum dot. 
Hence this time scale can be increased (and thus the required repetition rate 
decreased) by lowering the Fermi energy or having a flat band (thus reducing the 
Fermi velocity) or by applying a strong perpendicular magnetic field, 
which effectively localizes the reservoir modes. 

In a bigger picture, suppressing tunneling via the quantum Zeno effect and 
thereby effectively isolating the quantum dot from its environment is quite 
analogous to passive quantum error correcting schemes, such as the spin echo 
method. 
In contrast to active quantum error correcting schemes (typically involving 
redundancies), such passive schemes are based on interrupting the coherence 
between the quantum dot or qubit and its environment such that the constructive 
interference of the error amplitudes is turned into a (at least partially) 
destructive interference.
As an intuitive picture, these methods work well as long as the wave-packet 
associated with the quantum error is still ``lurking'' in the vicinity of the 
quantum dot or qubit.
Once they moved too far away, the impact of measurements or echo sequences is 
strongly reduced. 

\new{As an outlook, frequent measurements can also enable or accelerate 
quantum evolution, which is usually referred to as the anti-Zeno effect~\cite{kofmann2000a}.
As shown in the supplement, this effect can in principle also be realized 
in quantum dots, opening further windows for manipulation.} 

\paragraph{\new{Experimental scenarios}} 

In order to discuss the experimental relevance of our findings, let us insert 
typical experimental parameters.
For an effective mass of around $7~\%$ of the electron mass~\cite{fricke1996a} 
and a 2D electron 
density between $3$ and $6\times10^{11}\,\rm cm^{-2}$, we get Fermi energies 
between $10$ and $20~\rm meV$ 
and Fermi velocities between $2$ and $4\times10^5\,\rm m/s$. 
Assuming a typical level spacing in the quantum dot of around 
$50~\rm meV$~\citep{fricke1996a,warburton1998a}, 
the characteristic length scale of the ground state wave function is 
$\ell\approx5~\rm nm$. 
With a Fermi momentum of order $10^8\,\rm m^{-1}$, 
we have $k_{\rm F} \ell$ of order unity.  
Thus, we may estimate the crossover time $T_*$ by comparing the Fermi velocity 
with the length scale $\ell\approx5~\rm nm$ which yields rather short times 
of order $T_*=\ord(10~\rm fs)$. 
Unfortunately, although measurements can be performed quite fast 
(on the nano-second scale~\cite{kurzmann2019a}), this $T_*=\ord(10~\rm fs)$ 
is probably too short 
to observe the quantum Zeno effect in this set-up. 
Increasing this time scale $T_*$ by decreasing the Fermi energy 
(i.e., reducing the electron density in the reservoir) or by using 
materials with a flat band is possible in principle, but experimentally 
quite challenging. 

Thus, probably the most viable option is to apply a strong perpendicular
magnetic field $B$, say, of around $10~\rm T$,
such that the 2DEG is in the Quantum Hall 
regime~\footnote{In the Quantum Hall regime, small residual electric 
fields within
the 2DEG that originate e.g. from interface roughness
or electrostatic potential fluctuations introduced by the
QDs will result in a drift of the wave packet in crossed
$\f{E}$ and $\f{B}$ fields. The drift velocity 
$\f{v}_{\rm drift} = c(\f{E} \times \f{B})$
is, however, much lower than the Fermi velocity in the
original, field-free 2DEG. Thus, the problem of a temporally 
decreasing overlap is diminished.}
The resulting magnetic length of $\ell_B\approx 8~\rm nm$ 
is roughly of the same order as the length scale $\ell\approx5~\rm nm$ 
associated with 
the quantum dot and similarly the Landau level spacing of about 
$15~\rm meV$ 
is not too far away from the other energy scales. 
With aligning the energy levels accordingly, one can arrange resonant 
tunnel transitions between the discrete level of the quantum dot and one 
of the discrete Landau levels in the reservoir. 
In this case, the line of arguments sketched in the Introduction can be 
applied and thus the measurement repetition rate is set by the tunneling 
rate $\gamma$ between these two levels. 
Since the time associated scale $1/\gamma$ can be quite long (in the micro- 
to milli-second regime or even longer, which would also enable an electric 
readout~\cite{fujisawa2006a,flindt2009a}), 
a measurement time in the nano-second regime should be sufficient to observe 
the quantum Zeno effect. 

In order to enable such an observation, one has to make sure that the 
environmental decoherence (which can have basically the same effect as 
a measurement; for a review
of tunneling in a dissipative environment see e.g.~\cite{caldeira1983a}) 
is weak enough, i.e., the associated coherence time is 
longer than the time between measurements. 
A frequently discussed decoherence mechanism is the scattering of the 
electrons or holes in the reservoir at local impurities.
As long as these impurities 
(which can be characterized experimentally~\cite{kerski2021a}) 
are not too dense and thus far away from the dot, 
this mechanism is not relevant here because, as explained above, 
if the reservoir wave packet moved away from the dot far enough to see the 
impurity, it is already outside the region of applicability
of the quantum Zeno effect. 
Another potentially important mechanism stems from the Coulomb interaction 
between the electrons or holes or their interaction with phonons. 
In order to suppress this decoherence mechanism, the temperature should 
be low enough such that the available phase space and number of excitations 
is reduced. 

Also in this respect, working in a strong magnetic field can be
helpful: While electron-phonon scattering in a field-free
2DEG happens in two-dimensional $k$-space, in the Quantum Hall regime 
electron-phonon scattering is limited
to the one-dimensional (both in real-space and $k$-space) lines along 
which the electrons drifts. 
This reduced dimensionality strongly suppresses electron-phonon 
scattering and renders it negligible at milli-Kelvin temperatures. 
At these temperatures $T$, electron-electron scattering is the only 
relevant mechanism and follows a $T^2$ law~\cite{altshuler1982a} 
in a 2DEG in a strong magnetic field. 
Experimental
studies of the universality in the Quantum Hall regime~\cite{koch1992a}
allow one to estimate the associated time scale.












\acknowledgments

R.S.~acknowledges valuable discussions with S.~Popescu and W.G.~Unruh. 
Funded by the Deutsche Forschungsgemeinschaft (DFG, German Research Foundation) 
-- Project-ID 278162697-- SFB 1242. 

\bibliographystyle{unsrtab}
\bibliography{references}


\appendix

\section{Detector dynamics}\label{APP:detector}

To evaluate the detector occupation, we first note that in the interaction picture, 
the photon annihilation operator grows linearly in time 
\bea
\hat b(t) = \hat b(0) + \zeta t \hat n_{\rm d}(0)\,.
\ea

The actually emitted photon number of our toy model then becomes to second order perturbation theory
\bea
\expval{\hat b^\dagger \hat b}_t &=& \zeta^2 \int\limits_0^t dt_1\int\limits_0^t dt_2
\int\limits_{\rm F} d^2k\;|\gamma_\mathbf{k}|^2
\exp\left\{-\frac{\zeta^2}{2}(t_1-t_2)^2\right\}
\nn&&\times
(t-t_1)(t-t_2)\exp\{-i(\epsilon_{\rm d}-\epsilon_\mathbf{k})(t_1-t_2)\}\,,
\ea
which also reflects the crossover from a $t^4$ scaling for $t \ll T_*$ and $t \zeta \ll 1$ to a reduced -- at most $t^3$ -- scaling for $t \gg T_*$.
The initial delayed response can be understood as the detector has to react to the dynamics of the system first.
If the detector is used to reveal the Zeno suppression as a primary system, one should keep in mind that the detector delay also imposes a lower bound on the measurement time. 
For example, when the dot is initially filled, already the lowest order contributes and we would obtain $\expval{\hat b^\dagger \hat b}_t \approx \zeta^2 t^2$.
Thus, if $N_{\rm thr}$ denotes the minimum number of photons allowing to discriminate a charged dot from an empty dot, we obtain a lower bound on the measurement time $\zeta t \ge \sqrt{N_{\rm thr}}$.
To investigate whether one can simultaneously realize a Zeno-freezing of the dot while faithfully detecting it with the same primary detector, we can analytically perform the temporal integrals  in Eq.~\eqref{EQ:pocc_withdet}.
A quadratic temporal scaling is still possible but then puts tighter constraints on the $\gamma_k$.

The expectation value of a photon annihilation operator (relevant e.g. for homodyne measurements) would become instead
\bea
\expval{\hat b}_t &=& \zeta \int\limits_0^t dt_1\int\limits_0^t dt_2
\int\limits_{\rm F} d^2k\;|\gamma_\mathbf{k}|^2
\exp\left\{-\frac{\zeta^2}{2}(t_1-t_2)^2\right\}
\nn&&\times
(t-t_2)\exp\{-i(\epsilon_{\rm d}-\epsilon_\mathbf{k})(t_1-t_2)\}
\ea
and exhibits a crossover from a cubic to (at most) quadratic scaling.

Considering the effect of a finite detector frequency $\hat H_{\rm measure}\to \hat H_{\rm measure}+\omega_{\rm det} \hat b^\dagger \hat b$ implies an oscillatory behaviour of $\hat b(t)$ in the interaction picture and would eventually again lead to a linear long-term scaling in the Fermi golden rule regime.

\section{Measuring empty dots}\label{APP:empty}

If the detector would be sensitive to the empty state, i.e.,
\bea
\hat H_{\rm measure} = \hat a_{\rm d} \hat a_{\rm d}^\dagger \left(\ii \zeta \hat b^\dagger + {\rm h.c.}\right)
\ea
instead of~\eqref{EQ:pointer} in the main text, we would get in the interaction picture
\bea
\hat a_{\rm d}(t) &=& e^{-i\epsilon_{\rm d}t} \exp\left\{-(\zeta\hat b^\dagger-{\rm h.c.})t\right\} \hat a_{\rm d}(0)\,,\nn
\hat b(t) &=& \hat b(0) + \zeta t \hat a_{\rm d}(0) a_{\rm d}^\dagger(0)\,.
\ea
Since in the first line, we have only a flip in sign of $\zeta$ in comparison to~\eqref{EQ:intpic_ms} in the main text, the arguments raised there for the dot occupation prevail
in this case and we retain~\eqref{EQ:pocc_withdet} in the main text (assuming of course the same initial conditions).

This is different however for the detector occupation.
In this case, due to 
${\rm Tr}\left\{\hat b^\dagger(t) \hat b(t) \hat\rho_0\right\} = \zeta^2 t^2$, 
a detector signal is seen already for $\gamma_\mathbf{k}=0$ for an initially empty dot.
Up to second order, we get
\bea
\expval{\hat b^\dagger \hat b}_t &=& \zeta^2 t^2\\
&&+ \zeta^2 \int\limits_0^t dt_1\int\limits_0^t dt_2
\int\limits_{\rm F} d^2k\;|\gamma_\mathbf{k}|^2
\exp\left\{-\frac{\zeta^2}{2}(t_1-t_2)^2\right\}
\nn&&\times
[t_1 t_2 + t |t_1-t_2| - t^2]\exp\{-i(\epsilon_{\rm d}-\epsilon_\mathbf{k})(t_1-t_2)\}\,,\nonumber
\ea
where the temporal crossover of the second order contribution is much harder to observe as the first order contribution is dominant.
Thus, to see Zeno signatures in such a detector it would be more suitable to consider the reverse process of an electron tunnelling out of the dot.

\section{Zeno effect and reservoir characteristics}

To investigate how the simple two-level picture is influenced by a continuous reservoir, one may instead consider a finite-chain Hamiltonian 
(terms with a finite homogeneous on-site energy may be easily added)
\bea
\hat H_N^{\rm 1d} = \gamma \sum_{n=1}^{N-1} \left[\ket{n}\bra{n+1}+{\rm h.c.}\right]\,,
\ea
where $\gamma$ denotes the tunnelling rate and $N=2$ recovers the example in the introduction of the main text.
Since the model is homogeneous, an explicit diagonalization is possible via the transformation
\bea
\ket{n} = \sqrt{\frac{2}{N+1}} \sum_{k=1}^N \sin\left(\frac{\pi k n}{N+1}\right) \ket{E_k}\,,
\ea
and in terms of the new basis $\ket{E_k}$ the Hamiltonian becomes diagonal
\bea
\hat H_N^{\rm 1d}=\sum_{k=1}^N E_k \ket{E_k}\bra{E_k}\,,\;
E_k = -2\gamma\cos\left(\frac{\pi k}{N+1}\right)\,.
\ea
This facilitates the computation of the propagator and derived time-dependent quantities. 
For example, the probability of an initial excitation to remain at the first site becomes
\bea
P_{\rm stay}^{N,\rm 1d}(t) &=&  \abs{\bra{1} e^{-\ii \hat H_N^{\rm 1d} t} \ket{1}}^2\nn
&=& \abs{\sum_{k=1}^N e^{-\ii E_k t} \frac{2}{N+1} \sin^2\left(\frac{\pi k}{N+1}\right)}^2\,,
\ea 
which for $N=2$ just reduces to the $\cos^2(\gamma t)$ expression.
For an infinitely long chain we may represent the sum by an analytically solvable integral to obtain
\bea
P_{\rm stay}^{\infty,\rm 1d}(t) &=& \abs{\frac{{\cal J}_1(2\gamma t)}{\gamma t}}^2\,.
\ea

For higher-dimensional homogeneous lattices, the calculations can be linked to the one-dimensional case.
For example, in two-dimensions we can write the Hamiltonian as
\bea
\hat H_{N_x N_y}^{\rm 2d} &=& \gamma \sum_{n_x=1}^{N_x-1} \sum_{n_y=1}^{N_y} \left[\ket{n_x n_y}\bra{n_x+1, n_y} + {\rm h.c.}\right]\nn
&&+ \gamma \sum_{n_x=1}^{N_x} \sum_{n_y=1}^{N_y-1} \left[\ket{n_x n_y}\bra{n_x, n_y+1} + {\rm h.c.}\right]\nn
&=& \hat H_{N_x}^{\rm 1d} \otimes \f{1}_{N_y} + \f{1}_{N_x} \otimes \hat H_{N_y}^{\rm 1d}\,.
\ea
The trivial tensor product decomposition of the Hamiltonian then yields in $D$ dimensions
\bea
P_{\rm stay}^{\infty, D{\rm d}}(t) &=& \abs{\frac{{\cal J}_1(2\gamma t)}{\gamma t}}^{2D}\,.
\ea

However, one-dimensional chain representations remain also useful for generic reservoirs.
For example, starting from a representation where the reservoir (described by sites $2 \ldots N$) is diagonal, the Hamiltonian reads
\bea\label{EQ:hamrep1}
\hat H &=& \epsilon \ket{1}\bra{1} + \sum_{k=2}^N \epsilon_k \ket{k}\bra{k}\nn
&&+ \sum_{k=2}^N \left[t_k \ket{1}\bra{k} + t_k^* \ket{k}\bra{1}\right]\,,
\ea
where $\epsilon$ is the system energy, $\epsilon_k$ are the reservoir energies, and $t_k$ describe the system-reservoir tunnelling amplitudes.
We can always introduce a new basis in the reservoir only
\bea
\ket{k} = \sum_{q=2}^N u_{kq} \ket{\bar q}\,,
\ea
where $(u_{kq})$ is a unitary matrix with $u_{11}=1$ and $u_{1, k\ge 2}=u_{k\ge 2, 1} = 0$.
To further specify the transformation, we additionally impose the conditions
\bea\label{EQ:conditions}
\delta_{q2} &=& \sum_{k=2}^N \frac{t_k^*}{\sqrt{\sum_{\ell=2}^N \abs{t_\ell}^2}} u_{kq} \qquad:\qquad q\ge 2\,,\nn
\Omega_q \delta_{qq'} &=& \sum_{k=2}^N \epsilon_k u_{kq} u_{kq'}^*\qquad:\qquad q,q' \ge 3\,.
\ea
These conditions can be fulfilled by interpreting the $u_{kq}$ coefficients as eigenvector entries for an 
$(N-1)\times (N-1)$ hermitian diagonalization problem
\bea
M &=& \left(\f{1}-\f{v} \f{v}^\dagger\right)\left(\begin{array}{cccc}
\epsilon_2&0&\hdots&0\\
0&\ddots &\ddots &\vdots\\
\vdots & \ddots & \ddots & 0\\
0& \hdots &0&\epsilon_N\end{array}\right)
\left(\f{1}-\f{v} \f{v}^\dagger\right)\,,\nn
\f{v} &=& \frac{1}{\sqrt{\sum_{\ell=2}^N \abs{t_\ell}^2}} \left(\begin{array}{c}
t_2\\
\vdots\\
t_N
\end{array}\right)\,.
\ea
With conditions~\eqref{EQ:conditions}, the Hamiltonian is recast into 
\bea\label{EQ:hamrep2}
\hat H &=& \epsilon \ket{1}\bra{1} + \new{\gamma_{\rm RC}} \left[\ket{1}\bra{\bar 2}+\ket{\bar 2}\bra{1}\right] + \Omega_{\rm RC} \ket{\bar 2}\bra{\bar 2}\nn
&&+\sum_{k=3}^N \left[\tau_k \ket{\bar 2}\bra{\bar k} + \tau_k^* \ket{\bar k}\bra{\bar 2}\right]
+\sum_{k=3}^N \Omega_k \ket{\bar k}\bra{\bar k}\,,\nn
\new{\gamma_{\rm RC}} &=& \sqrt{\sum_{\ell=2}^N \abs{t_\ell}^2}\,,\qquad
\Omega_{\rm RC} = \sum_{k=2}^N \epsilon_k \abs{u_{k2}}^2 = \sum_{k=2}^N \epsilon_k \frac{\abs{t_k}^2}{\new{\gamma_{\rm RC}^2}}\,,\nn
\tau_k &=& \sum_{q=2}^N \epsilon_q u_{q2} u_{qk}^* = \sum_{q=2}^N \epsilon_q \frac{t_q}{\new{\gamma_{\rm RC}}} u_{qk}^*\,. 
\ea
In contrast to the starting Hamiltonian~\eqref{EQ:hamrep1}, the original system $\ket{1}$ is now first coupled to a collective coordinate $\ket{\bar 2}$ with energy $\Omega_{\rm RC}$ via the amplitude $\new{\gamma_{\rm RC}}$.
The collective coordinate $\ket{\bar 2}$ is then in turn coupled to the residual reservoir just as in the original Hamiltonian the mode $\ket{1}$ was coupled to its reservoir.
Thus, since this Hamiltonian is structurally equivalent to our initial one, one can apply the procedure recursively.
Eventually, this will map the reservoir to a (non-homogeneous) chain.

To consider reservoirs of infinite size we introduce the original spectral coupling density
$\Gamma(\omega) \equiv 2\pi \sum_k \abs{t_k}^2 \delta(\omega-\epsilon_k)$
as a continuous (not necessarily differentiable) function.
The tunnelling amplitude and energy of the collective coordinate maintain their meaning and can be written as
\bea\label{EQ:RCparams}
\new{\gamma_{\rm RC}^2} = \frac{1}{2\pi} \int \Gamma(\omega) d\omega\,,\;\;\;
\Omega_{\rm RC} = \frac{1}{2\pi\new{\gamma_{\rm RC}^2}} \int \omega \Gamma(\omega) d\omega\,.
\ea
The residual (new) spectral coupling density $\bar\Gamma(\omega)\equiv 2\pi \sum_k \abs{\tau_k}^2 \delta(\omega-\Omega_k)$ is also a function of the old spectral coupling density
\bea\label{EQ:mapping}
\bar\Gamma(\omega) &=&  \frac{4\new{\gamma_{\rm RC}^2} \Gamma(\omega)}{\left[\frac{1}{\pi} {\cal P} \int \frac{\Gamma(\omega')}{\omega'-\omega} d\omega'\right]^2 + \left[\Gamma(\omega)\right]^2}\,.
\ea
This formula is more complicated to obtain but can for example be found by solving the Heisenberg equations of motion for $\ket{1}\bra{1}$ according to Eq.~\eqref{EQ:hamrep1} and Eq.~\eqref{EQ:hamrep2} in analogy to Ref.~\cite{nazir2019a}.
Shortcutting this, one may also express the old spectral \new{coupling} density by
\bea
\Gamma(\omega)&=& \int \bra{1} e^{+\ii \hat H_B\tau} \hat H_I e^{-\ii \hat H_B \tau} \hat H_I \ket{1} e^{+\ii\omega\tau} d\tau\nn
&=& \new{\gamma_{\rm RC}^2} \int \bra{\bar 2} e^{-\ii \hat H_B \tau} \ket{\bar 2} e^{+\ii\omega\tau} d\tau\,,
\ea
where $\hat H_I = \hat h+\hat h^\dagger$ with $\hat h = \sum_k t_k \ket{1}\bra{k} = \new{\gamma_{\rm RC}} \ket{1}\bra{\bar 2}$ and
$\hat H_B = \sum_k \epsilon_k \ket{k}\bra{k} = \Omega_{\rm RC} \ket{\bar 2}\bra{\bar 2} + \sum_k \left[\tau_k \ket{\bar 2}\bra{\bar k} + \tau_k^* \ket{\bar k}\bra{\bar 2}\right]
+\sum_k \Omega_k \ket{\bar k}\bra{\bar k}$.
Inserting the operators in the original basis recovers the definition of the old spectral \new{coupling} density, whereas inserting the new basis yields an expression for the residual (new) spectral \new{coupling} density.
Defining $P_{22}(\tau) = \bra{\bar 2} e^{-\ii \hat H_B \tau} \ket{\bar 2}$, one obtains for its Laplace transform $p_{22}(z)=\int P_{22}(\tau) e^{-z \tau} d\tau$ the expression
\bea
p_{22}(z) = \frac{1}{z+\ii\Omega_{\rm RC}+\frac{1}{2\pi} \int \frac{\bar\Gamma(\omega')}{z+\ii\omega'} d\omega'}\,.
\ea
For a function obeying $f(t)=f^*(-t)$, its Laplace transform $F(z)=\int_0^\infty f(t) e^{-z t} dt$ provides information about its even and odd Fourier transforms
\bea
\lim_{\epsilon\to 0^+} 2 \Re F(\epsilon-\ii\omega) &=& \int f(t) e^{+\ii\omega t} dt = \tilde f(\omega)\,,\nn
\lim_{\epsilon\to 0^+} 2\ii \Im F(\epsilon-\ii\omega) &=& \int f(t) {\rm sgn}(t) e^{+\ii\omega t} dt\nn
&=& \frac{\ii}{\pi} {\cal P} \int \frac{\tilde f(\omega')}{\omega-\omega'} d\omega'\,.
\ea
From the first line above, it follows that $\Gamma(\omega)=\new{\gamma_{\rm RC}^2} \lim\limits_{\epsilon\to 0}[p_{22}(\epsilon-\ii\omega)+{\rm h.c.}]$, and we obtain a relation between the old and the residual (new) spectral coupling densities
\bea
\Gamma(\omega) = \frac{\new{\gamma_{\rm RC}^2} \bar\Gamma(\omega)}{\left[\omega-\Omega_{\rm RC}-{\cal P}\int \frac{\bar\Gamma(\omega')}{\omega-\omega'} \frac{d\omega'}{2\pi}\right]^2 + \left[\frac{\bar\Gamma(\omega)}{2}\right]^2}\,,
\ea
which precisely corresponds to the inverse reaction-coordinate mapping, cf. Eq.~(12) in Ref.~\cite{martensen2019a}.
Considering the imaginary part of the Laplace transform instead allows -- together with the relation above -- to solve also for the residual spectral \new{coupling} density $\bar\Gamma(\omega)$ in dependence of the old one, yielding precisely Eq.~\eqref{EQ:mapping}, which corresponds to the particle mapping for bosons or fermions, cf. Eqns.~(11) and~(14) of Ref.~\cite{nazir2019a}.

To relate this to the Zeno time, we note that for a single resonant level model, the Zeno transition time is directly given by the spectral \new{coupling} density 
\bea
T_* = \frac{2\pi\Gamma(\epsilon) f(\epsilon)}{\int \Gamma(\omega) f(\omega) d\omega} \to \frac{2\pi\Gamma(\epsilon)}{\int \Gamma(\omega) d\omega}\,,
\ea
where \new{$f(\omega)\equiv[e^{\beta(\omega-\mu)}+1]^{-1}$ denotes the Fermi function with inverse temperature $\beta$ and chemical potential $\mu$ and} the limit holds for a completely filled reservoir (high chemical potentials and low temperatures).

If we consider a Lorentzian spectral \new{coupling} density 
\bea
\Gamma(\omega) = \frac{\Gamma \delta^2}{(\omega-\epsilon_B)^2+\delta^2}\,,
\ea
this would imply $\bar\Gamma(\omega)=2\delta$ and for the Zeno time
$T_* = \frac{2\delta}{(\epsilon-\epsilon_B)^2+\delta^2}$,
which can be made very large when $\epsilon\approx \epsilon_B$ and $\delta \to 0$.
Indeed, considering the limit $\delta\to 0$ and $\Gamma\to\infty$ such that $\Gamma \delta$ remains constant, we obtain an effective two-level system
\bea
\hat H_S = \epsilon \ket{1}\bra{1} + \epsilon_B \ket{\bar 2}\bra{\bar 2} + \sqrt{\frac{\Gamma \delta}{2}} (\ket{1}\bra{\bar 2} + {\rm H.c.})
\ea
that is decoupled from the residual reservoir.

Thus, generic reservoirs may reproduce the two-level picture when they support a highly peaked spectral \new{coupling} density.

\new{

\section{Zeno and Anti-Zeno dynamics}

So far, we discussed the standard Zeno effect, i.e., how frequent measurements slow down the quantum evolution compared to the unperturbed one. 
Depending on the level configurations, one may however also observe an acceleration of the decay dynamics, termed anti-Zeno~\cite{kofmann2000a} or inverse Zeno~\cite{facchi2001a} effect.
Generically, one may express the unperturbed expectation value of a positive definite operator $\hat{O}$ with initial expectation value $\expval{\hat O}_0=1$ for a quantum system subject to Hamiltonian $\hat H$ as $P_{\rm stay}(t) = \tr\{\hat O e^{-\ii \hat H t} \rho_0 e^{+\ii \hat H t}\} \equiv e^{-R(t) \cdot t}$.
The effective rate $R(t)$ is in general different from its Fermi Golden Rule value $R_{\rm FG} = {\rm const}$.
If projective measurements are performed at time intervals $\tau$, the remainder probability after $n$ measurements with $t=n\tau$ will be given by $[e^{-R(\tau) \tau}]^n = e^{-R(\tau) t}$ as compared to a measurement-free (Fermi Golden Rule) evolution $e^{-R_{\rm FG} t}$ for large times $t$.
Thus, when $R(\tau)>R_{\rm FG}$ one may actually accelerate the decay of the initial state by measurement (anti-Zeno effect), 
whereas when $R(\tau) < R_{\rm FG}$ the decay may be slowed down (conventional Zeno effect).
The relevant effective decay rate can be obtained via 
\begin{align}\label{EQ:decay_rate}
R(\tau) = \frac{-\ln P_{\rm stay}(\tau)}{\tau}\,.
\end{align}
This is already observable for the previously discussed reservoir with a Lorentzian spectral coupling density. 
To connect with the main text, we discuss this for the single resonant level subject to a reservoir with the Gaussian model below.

\section{Example: Single resonant level with exponential spectral function}

The single resonant level coupled to a reservoir is described by the fermionic total Hamiltonian
\bea\label{EQ:srl}
\hat H = \epsilon_{\rm d} \hat{n}_d + \sum_k \epsilon_k \hat c_k^\dagger \hat c_k + \sum_k \left[t_k \hat d^\dagger \hat c_k + t_k^* \hat c_k^\dagger \hat d\right]\,,
\ea
where the first term denotes the dot Hamiltonian with $\hat n_d = \hat d^\dagger \hat d$ and on-site energy $\epsilon_{\rm d}$, the second term the reservoir with energies $\epsilon_k$, and the last term the interaction, for continuum reservoirs described by the spectral coupling density $\Gamma(\omega) = 2\pi \sum_k \abs{t_k}^2 \delta(\omega-\epsilon_k)$.
When we consider the Gaussian model~\eqref{Gaussian} from the main text, the spectral coupling density in absence of a magnetic field assumes an exponentially decaying form
\begin{align}\label{EQ:specdens_gauss}
\Gamma(\omega) = \Gamma_0 e^{-\lambda\omega} \Theta(\omega)\,,
\end{align} 
where $\Gamma_0=m \gamma_0^2$ and $\lambda = 2 m \ell^2$.
Thus, for small dot size $\ell$, the spectral coupling density decays very slowly and can be considered nearly flat. 
The Zeno transition time from the main text becomes $T_* = 2\pi\lambda e^{-\lambda \epsilon_{\rm d}} \Theta(\epsilon_{\rm d})$.
In contrast, when we turn on the magnetic field, the levels in the reservoir immediately obtain a finite splitting (Landau levels), 
which increases linearly with the field strength.

We can address the dot occupation dynamics with several methods, which we make explicit for an initially empty dot $\expval{\hat d^\dagger \hat d}_0 = 0$ and a thermally occupied reservoir $\expval{\hat c_k^\dagger \hat c_k}_0 = f(\epsilon_k)$ with Fermi function $f(\omega)=[e^{\beta(\omega-\mu)}+1]^{-1}$:

First, the system dynamics can be solved exactly by solving the Heisenberg equations of motion~\cite{landi2022a} for the fermionic operators $\hat d$ and $\hat c_k$.
After some algebra this yields for the dot occupation
\bea\label{EQ:dotocc_ex}
\expval{\hat n_{\rm d}}_t^{\rm ex} &=& \frac{1}{2\pi} \int \Gamma(\omega) f(\omega) \abs{g(\omega,t)}^2 d\omega\,,\\
\int\limits_0^\infty g(\omega,t) e^{-s t} dt &=&  \frac{1}{\left(s+\ii\omega\right)\left(s+\ii\epsilon_{\rm d}+\frac{1}{2\pi}\int\frac{\Gamma(\omega')}{s+\ii\omega'} d\omega'\right)}\,,\nonumber
\ea
which is still hard to evaluate due to the required inversion of a Laplace transform.
At low temperatures, the Fermi function becomes a step function $f(\omega)\to\Theta(\mu-\omega)$.

Second, we can use the second-order perturbative solution analogous to the discussion in the main text
\bea\label{EQ:dotocc_so}
\expval{\hat n_{\rm d}}_t^{\rm so} = \frac{t^2}{2\pi} \int \Gamma(\omega) f(\omega) {\rm sinc}^2\left[(\omega-\epsilon_{\rm d})\frac{t}{2}\right] d\omega\,,
\ea
which is already much simpler to evaluate.
The above formula has the short-term regimes $\expval{\hat n_{\rm d}}_t \approx \frac{t^2}{2\pi} \int \Gamma(\omega) f(\omega)d\omega$
and Fermi Golden Rule regimes $\expval{\hat n_{\rm d}}_t \approx t \Gamma(\epsilon_{\rm d}) f(\epsilon_{\rm d})$.
In regimes where the perturbative treatment is valid (i.e., where $\expval{\hat n_{\rm d}}_t$ is small, the effective decay rate~\eqref{EQ:decay_rate} can be further simplified
\bea\label{EQ:effrate_simp}
R(t) &=& \frac{-\ln[1-\expval{\hat n_{\rm d}}_t]}{t} \approx \frac{\expval{\hat n_d}_t^{\rm so}}{t}\nn
&=& \int \Gamma(\omega) f(\omega) \frac{t}{2\pi} {\rm sinc}^2\left[(\omega-\epsilon_{\rm d})\frac{t}{2}\right]d\omega\,,
\ea
which generically vanishes at small times (Zeno regime) and assumes the Fermi Golden rule rate $R_{\rm FG} = \Gamma(\epsilon_{\rm d}) f(\epsilon_{\rm d})$ at large times.

Third, we can perform the mapping to an effective two-level system~\eqref{EQ:RCparams} coupled to a residual reservoir with modified spectral coupling density~\eqref{EQ:mapping}
(also called spectral function).
If the latter is small, we can (for the initial times we are interested in) neglect the residual spectral coupling density and consider only the effective two-level dynamics
\bea\label{EQ:dotocc_2l}
\expval{\hat n_{\rm d}}_t^{\rm 2L} &=& \frac{4\gamma_{\rm RC}^2 \sin^2 \left[\frac{t}{2} \sqrt{4\gamma_{\rm RC}^2+(\epsilon_{\rm d}-\Omega_{\rm RC})^2}\right]}{4\gamma_{\rm RC}^2+(\epsilon_{\rm d}-\Omega_{\rm RC})^2}\,.
\ea
For the spectral coupling density~\eqref{EQ:specdens_gauss}, the mapping relation~\eqref{EQ:RCparams} yield for the effective two-level system the explicit parameters 
\bea\label{EQ:mapped2ls}
\gamma_{\rm RC} = \sqrt{\frac{\Gamma_0}{2\pi\lambda}}\,,\qquad
\Omega_{\rm RC} = \frac{1}{\lambda}\,,
\ea
whereas the residual spectral coupling density must be computed numerically via~\eqref{EQ:mapping}.

The results are depicted in Fig.~\ref{FIG:dotocc}.
\begin{figure*}
\begin{tabular}{ccc}
\includegraphics[width=0.4\textwidth,clip=true]{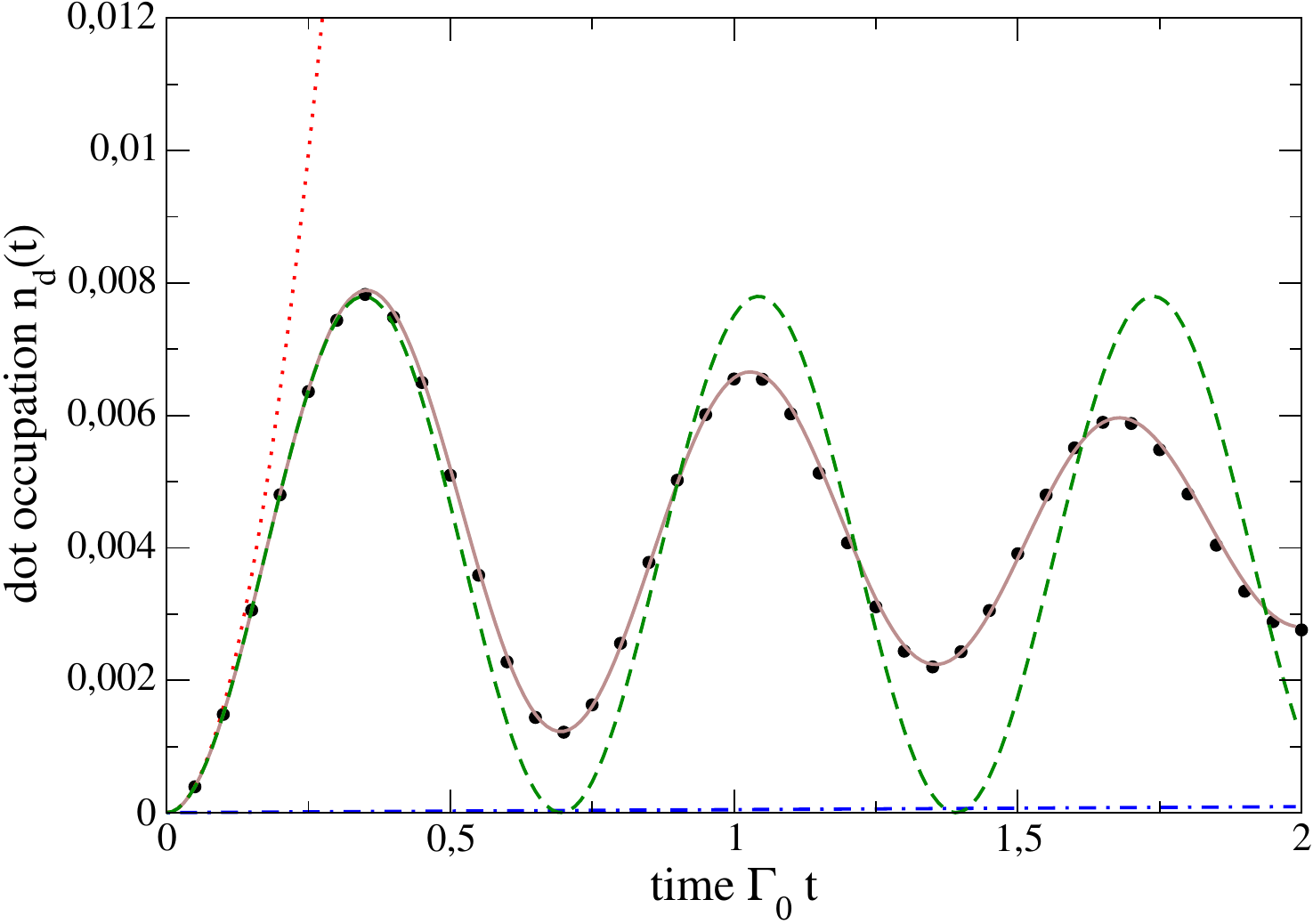} & \includegraphics[width=0.4\textwidth,clip=true]{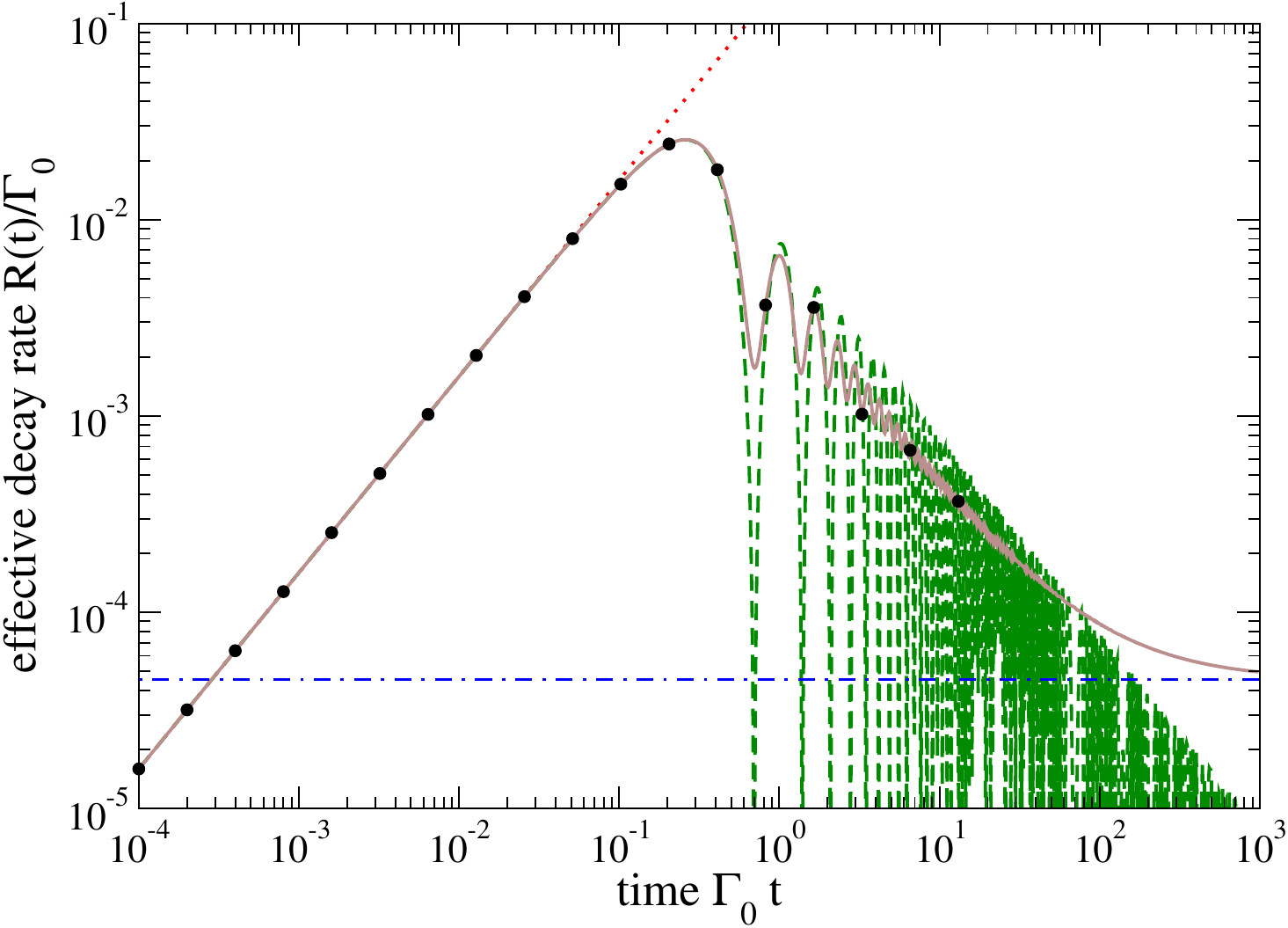} & \includegraphics[width=0.05\textwidth,clip=true]{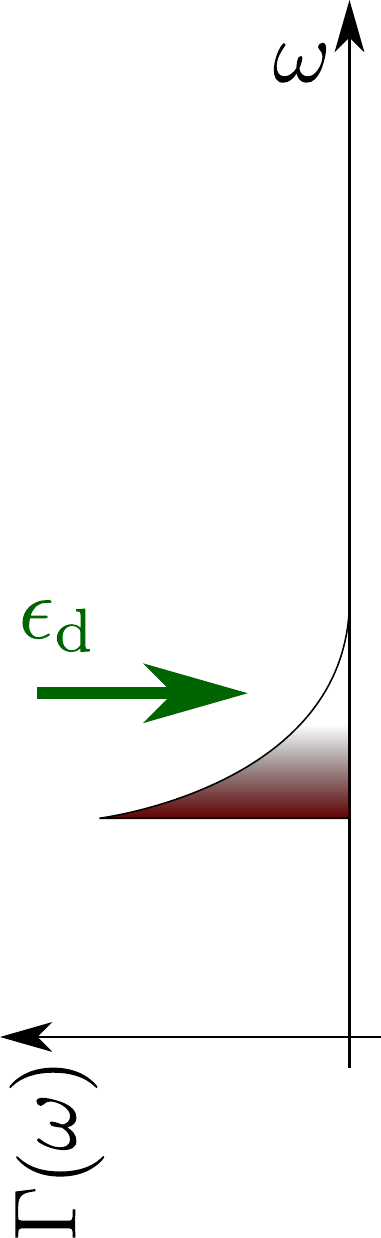}\\
\includegraphics[width=0.4\textwidth,clip=true]{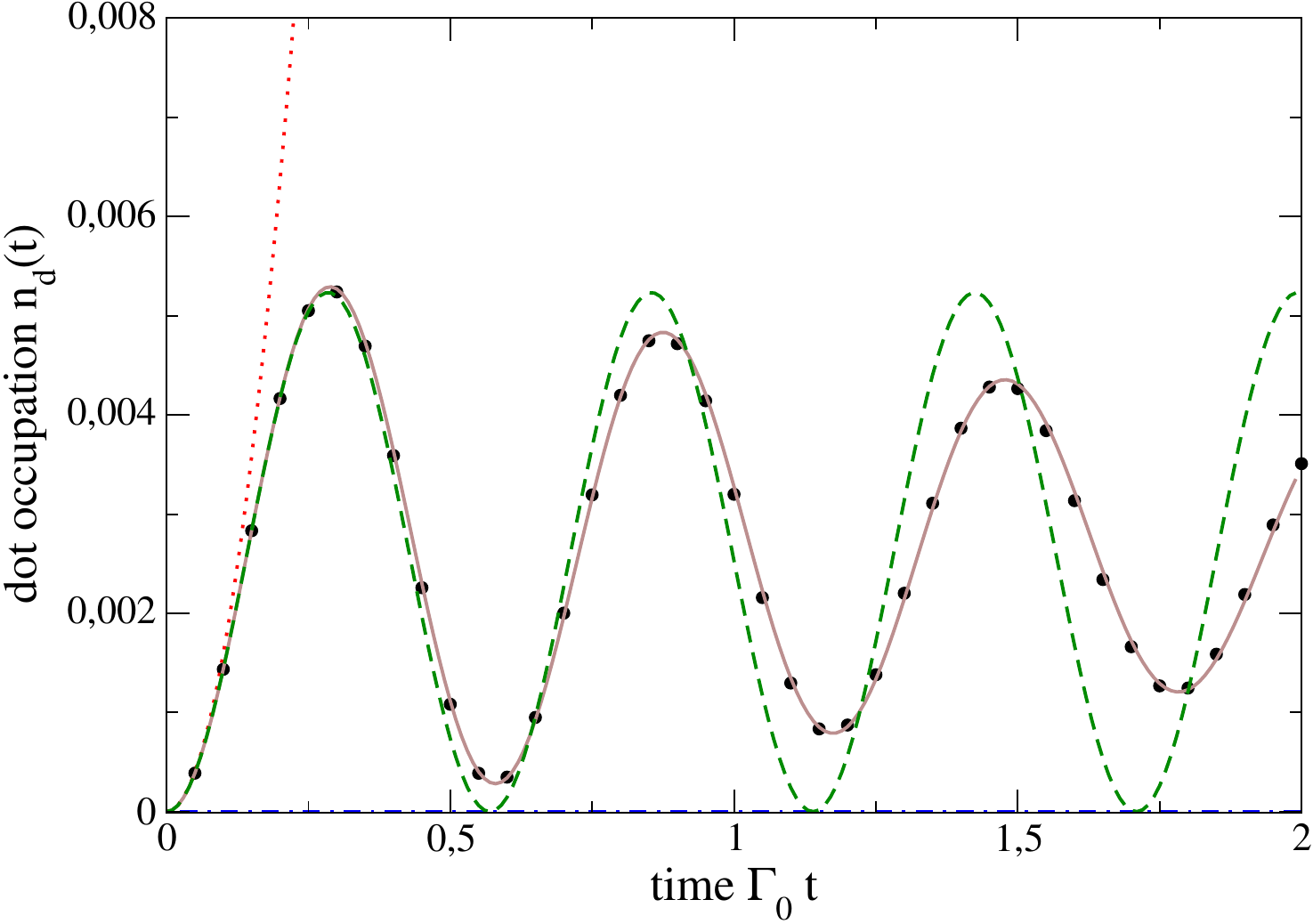} & \includegraphics[width=0.4\textwidth,clip=true]{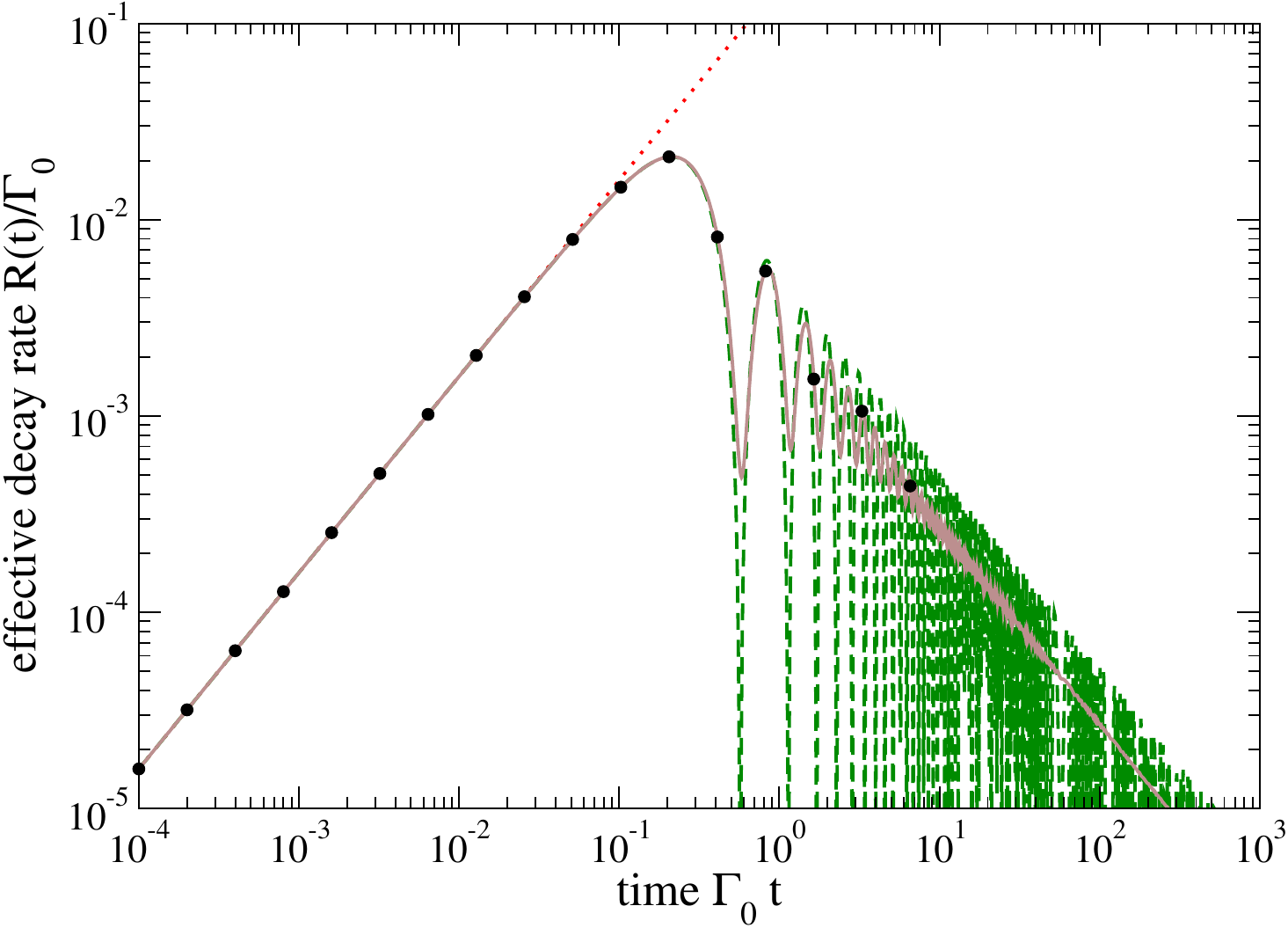} & \includegraphics[width=0.05\textwidth,clip=true]{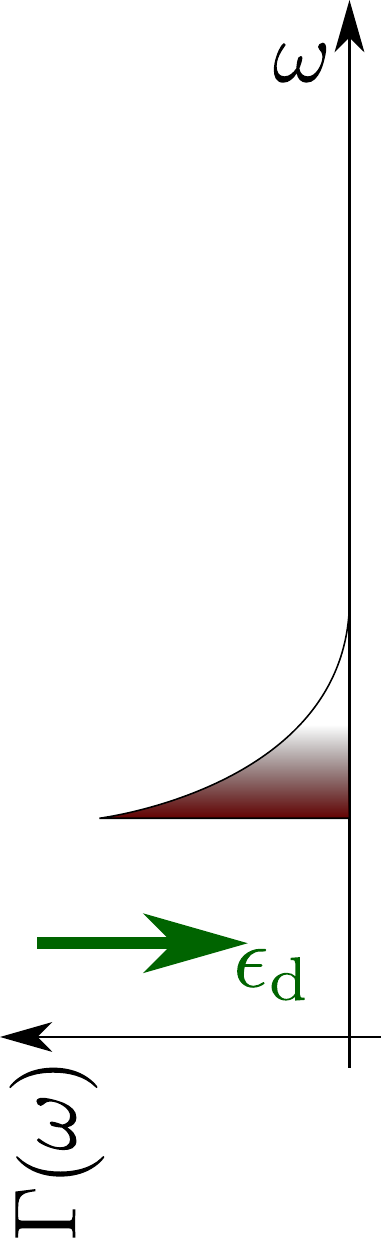}\\
\includegraphics[width=0.4\textwidth,clip=true]{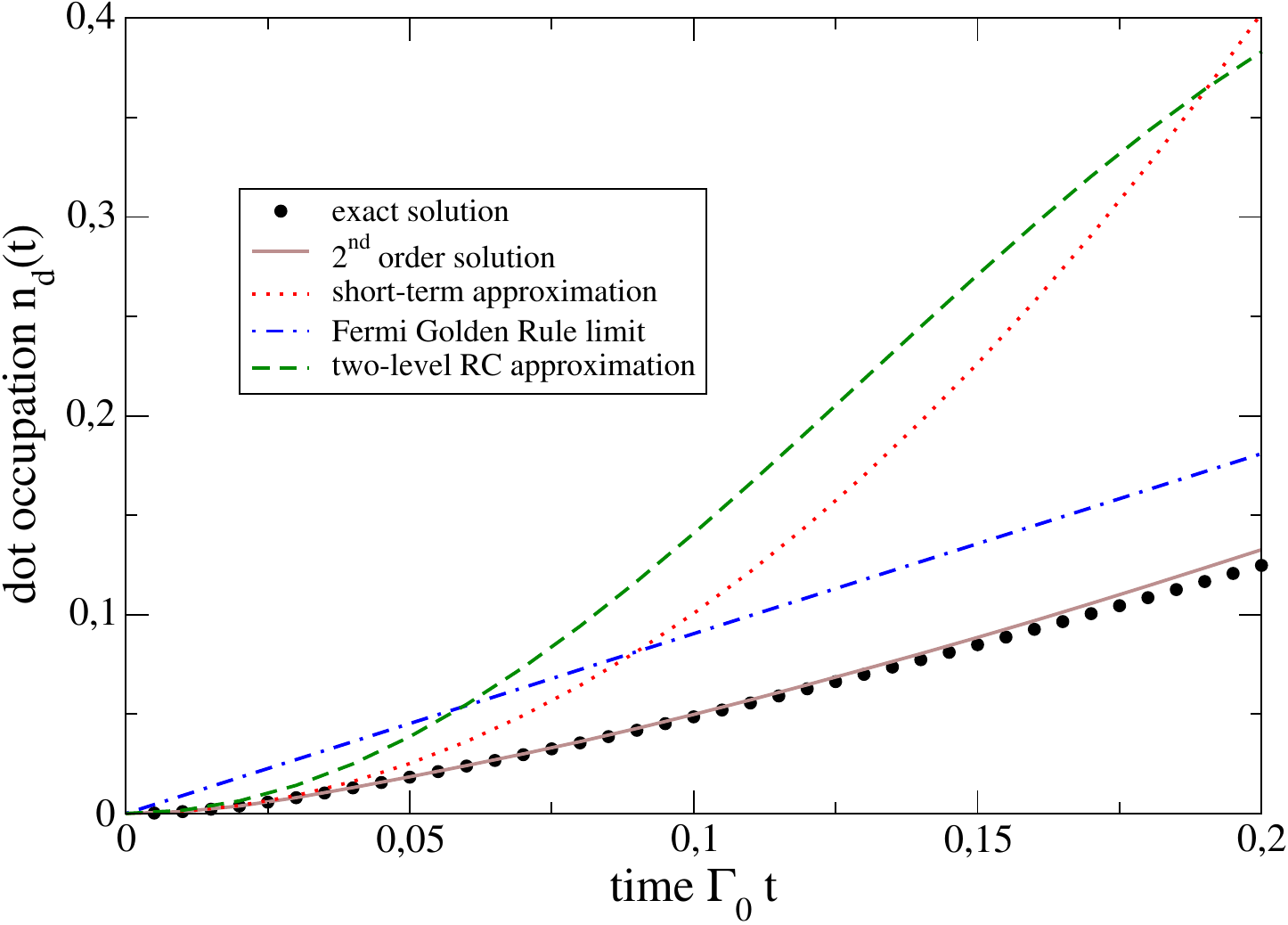} & \includegraphics[width=0.4\textwidth,clip=true]{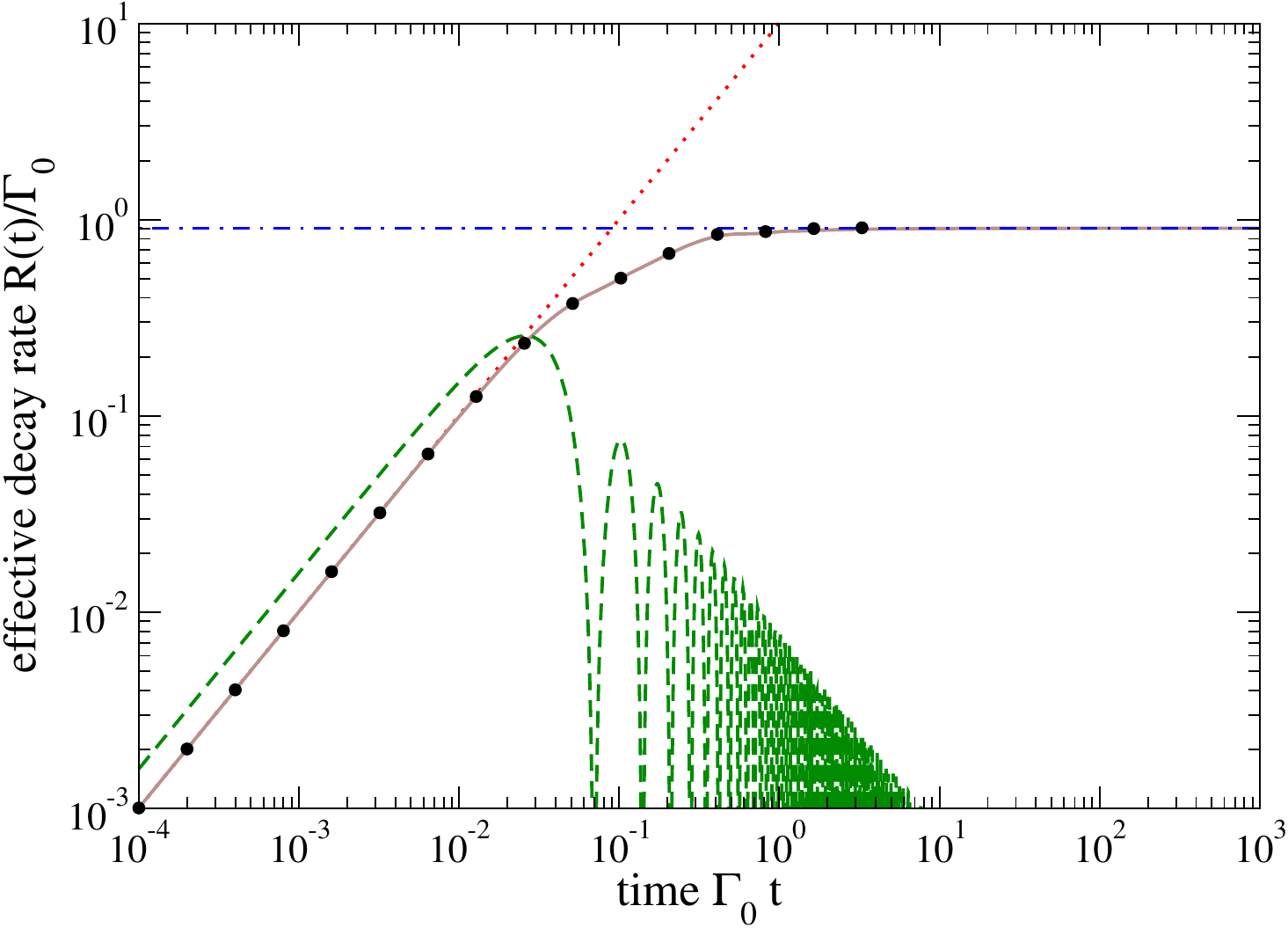} & \includegraphics[width=0.05\textwidth,clip=true]{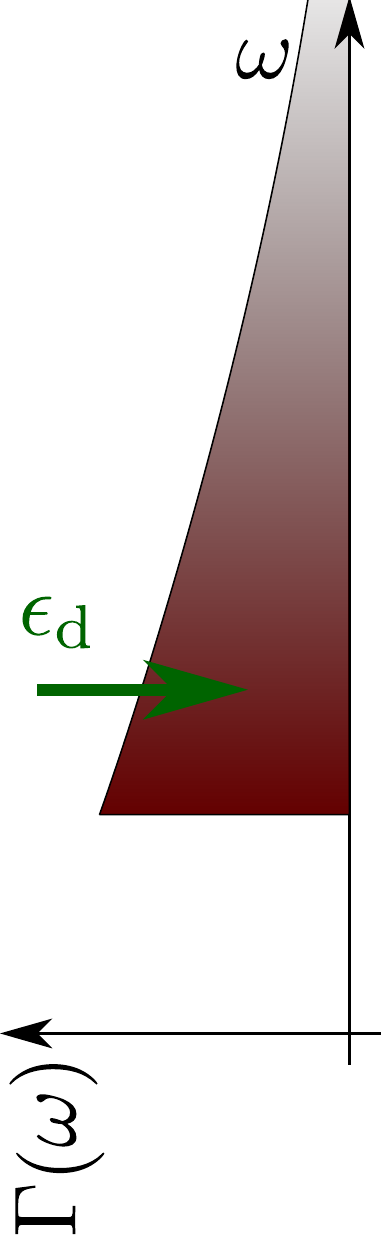}
\end{tabular}
\caption{\label{FIG:dotocc}
Dot occupation vs. time (left column) and effective decay rate (middle column) for a dot weakly coupled ($\Gamma_0 = 0.1\abs{\epsilon_{\rm d}}$) to a low-temperature and high-potential ($\beta \abs{\epsilon_{\rm d}} = 100$ and $\mu = 10 \abs{\epsilon_{\rm d}}$) reservoir described by spectral coupling density~\eqref{EQ:specdens_gauss} for different configurations (sketched in the right column).
In the top row, the level is inside the reservoir window with $\epsilon_{\rm d}>0$ and quickly decaying spectral coupling density $\lambda=10/\epsilon_{\rm d}$, 
in the middle row, the level is below the reservoir window with $\epsilon_{\rm d}<0$ and quickly decaying spectral coupling density $\lambda=10/\abs{\epsilon_{\rm d}}$, and
in the bottom row, the level inside the reservoir window with $\epsilon_{\rm d}>0$ and slowly decaying spectral coupling density $\lambda=0.1/\epsilon_{\rm d}$.
Black symbols denote the exact solution~\eqref{EQ:dotocc_ex} and the full effective rate $-t^{-1}\ln(1-\expval{n_{\rm d}}_t^{\rm ex})$, respectively, 
solid brown curves the second order solution~\eqref{EQ:dotocc_so} and the approximate effective decay rate~\eqref{EQ:effrate_simp}, respectively.
Dotted red and dash-dotted blue curves denote the short-term and Fermi Golden Rule approximations to the second order perturbative solution, respectively.
Finally, the dashed green curves depict the evolution due to the isolated effective two-level system~\eqref{EQ:dotocc_2l} with parameters chosen according to~\eqref{EQ:mapped2ls} and
the corresponding approximate effective decay rate~\eqref{EQ:effrate_simp}, respectively.
}
\end{figure*}

The first row displays the situation of a dot level inside the reservoir spectral window, which however has a quickly decaying spectral coupling density (top right sketch), such that
the effective two-level picture (dashed green curves) applies well.
For this off-resonant case, one finds a transition from a Zeno-regime to an anti-Zeno regime (where the effective decay rate is larger than the Fermi Golden Rule value, top right plot) before the Fermi Golden Rule regime is reached for long times.
Thus, the discussion from the main text is not applicable for this special case.
However, since the two-level picture applies to this regime, we can estimate the time range within which anti-Zeno effects are observable by equating the rates according to the two-level picture with the Fermi Golden Rule result
$\expval{n_{\rm d}}_{t^*}^{\rm 2L}/t^* = \Gamma(\epsilon_{\rm d}) f(\epsilon_{\rm d})$.

In the second row, the dot level is outside the reservoir spectral window, such that the Fermi Golden Rule result would forbid the dot to be loaded (middle left plot), 
and the corresponding rate just vanishes (middle right plot).
Nevertheless, loading is possible at finite times beyond leading order.
Projective measurements would manifest these virtual processes, such that there is no Zeno regime but only an anti-Zeno regime in this case.
Another way to look at this behaviour is that for small times, the energy-time uncertainty relation does not allow to resolve the energy mismatch between system and reservoir.

The bottom row depicts the situation discussed in the main text, where the spectral coupling density of the reservoir decays slowly. 
Although the effective two-level picture does not apply well (the residual spectral coupling density~\eqref{EQ:mapping} is not negligible), the maximum height of the oscillations demonstrates that an anti-Zeno effect is out of reach for this scenario. 
One finds a direct transition between Zeno and Fermi Golden Rule regimes, and the Zeno time $T^*$ is is a useful diagnostic measure.
}

\end{document}